\documentclass[twocolumn,showpacs,aps]{revtex4}

\usepackage{graphicx}

\newcommand{\wfmfig}[5]{%
\begin{figure}
\includegraphics[width=#3]{#2}
\caption[\protect{ #4}]{\protect\label{fg:#1} #5}%
\end{figure}
}

\begin{document}

\title{Variation with mass of  $\boldmath{B(E3; 0_1^+ \rightarrow 3_1^-)}$
transition rates in $A=124-134$ even-mass xenon nuclei}

\author{W.F.~Mueller$^1$\footnote{Present address: CANBERRA, 800 Research Parkway, Meriden, CT 06450.},
M.P.~Carpenter$^2$,
J.A.~Church$^{1,3}$\footnote{Present address: Lawrence Livermore
  National Laboratory, P.O.~Box 808, L-414, Livermore, CA 94550.} 
D.C.~Dinca$^{1,3}$\footnote{Present address: American Science and
  Engineering, Inc., 829 Middlesex Turnpike, Billerica, MA 01821.},
A.~Gade$^1$,
T.~Glasmacher$^{1,3}$,
D.T.~Henderson$^2$,
Z.~Hu$^1$\footnote{Present address: Philips Medical Systems, 595 Miner Rd, Cleveland, OH 44143.},
R.V.F.~Janssens$^2$,
A.F.~Lisetskiy$^1$\footnote{Present address: Theory Department,
  Planckstr. 1, 64291 Darmstadt, Germany.},
C.J.~Lister,$^2$,
E.F.~Moore$^2$,
T.O.~Pennington$^2$,
B.C.~Perry$^{1,3}$,
I.~Wiedenh\"{o}ver$^4$,
K.L.~Yurkewicz$^{1,3}$\footnote{Present address: Fermilab, MS 206(WH
  1E), Batavia IL 60510.},
V.G. Zelevinsky$^{1,3}$, and
H.~Zwahlen$^{1,3}$}

\affiliation
{$^1$National Superconducting Cyclotron Laboratory,
     Michigan State University, East Lansing, MI 48824-1321, USA}
\affiliation
{$^2$Physics Division, Argonne National Laboratory, Argonne, IL 60439, USA}
\affiliation
{$^3$Department of Physics and Astronomy,
     Michigan State University, East Lansing, MI 48824, USA}
\affiliation
{$^4$Department of Physics,
     Florida State University, Tallahassee, FL 32306, USA}

\date{\today}

\begin{abstract}
$B(E3; 0_1^+ \rightarrow 3_1^-)$ transition matrix elements have
been measured for even-mass $^{124-134}$Xe nuclei using
sub-barrier Coulomb excitation in inverse kinematics. The trends
in energy $E(3^-)$ and $B(E3; 0_1^+ \rightarrow 3_1^-)$ excitation
strengths are well reproduced using phenomenological models based
on a strong coupling picture with a soft quadrupole mode and an 
increasing occupation of the intruder $h_{11/2}$ orbital.
\end{abstract}

\pacs{21.10.Ky, 21.10.Re, 21.60.Ev, 21.60.Jz, 23.20.Lv, 27.60.+j, 29.40.Gx}
\keywords{}

\maketitle

Xenon isotopes with mass numbers from $A=124$ to 134 are located
in a transitional region of nuclei where their low-spin structure
indicates an evolution from a weakly-deformed, $\gamma$-soft rotor
to a vibrator. The positive-parity states of xenon nuclei in this
region have been described within the framework of algebraic
models~\cite{Cas85,Zel90,Neu96,Gad00,Wer01,Pan96}, of triaxial
rotor-vibrator models~\cite{Mey98}, and using a pair-truncated
shell-model description~\cite{Yos04}.  However, the low-spin,
negative-parity structures have not been extensively investigated.

The properties of such negative-parity states can provide
important information on the octupole degree of freedom. General
features of collective octupole motion are still not well
understood \cite{Met95}. Specifically, the excitation energy and,
more importantly, the $B(E3; 0_1^+ \rightarrow 3_1^-)=B(E3
\uparrow)$ transition probabilities for $3^-$ states are
signatures of either octupole vibrational strength or static
octupole deformation~\cite{Spe90,Naz84}.

While some information is available on the $B(E3 \uparrow)$
strength of nuclei in this mass region such as for
Sn~\cite{Jon81}, Te~\cite{Mat75}, and Ba~\cite{Bur85}, the only
data on octupole collectivity in xenon isotopes stems from an
inelastic proton-scattering measurement on
$^{136}$Xe~\cite{Sen72,Kib02}. The combination of soft quadrupole 
and octupole modes can make such nuclei as $^{129}$Xe good
candidates for the search of enhancement of the nuclear Schiff
moment and atomic electric dipole moment \cite{Fla03}. Here, we
present the first determination of the $B(E3 \uparrow)$ rates in
the chain of even-mass $^{124-134}$Xe isotopes. These results are
obtained under identical experimental conditions and compared to a
global model \cite{Met95} reproducing the trends of the $B(E3
\uparrow)$ strengths across the nuclear chart.

The xenon beams were delivered by the {\sc atlas} accelerator at
Argonne National Laboratory.  The beam energies for the
$A=124-134$ even xenon isotopes were 555, 556, 553, 563, 571, and
579~MeV, respectively. These beams were directed to the
experimental setup (see Fig.~\ref{fg:expsetup}) where they
impinged on a $^{58}$Ni target with 1.1~mg/cm$^2$ thickness and an
enrichment of $\geq$95\%. Scattered xenon nuclei were detected in
a large area, four quadrant, parallel-plate avalanche counter
({\sc ppac}). The entrance window was located 21.5~cm downstream
from the target.  The {\sc ppac} size was such that it covered
laboratory scattering angles between 9$^{\circ}$ and 40$^\circ$.
The maximum laboratory scattering angle for the xenon nuclei was
about 27$^\circ$. Hence, it was possible to cover the most
relevant scattering angular ranges.

\wfmfig{expsetup}{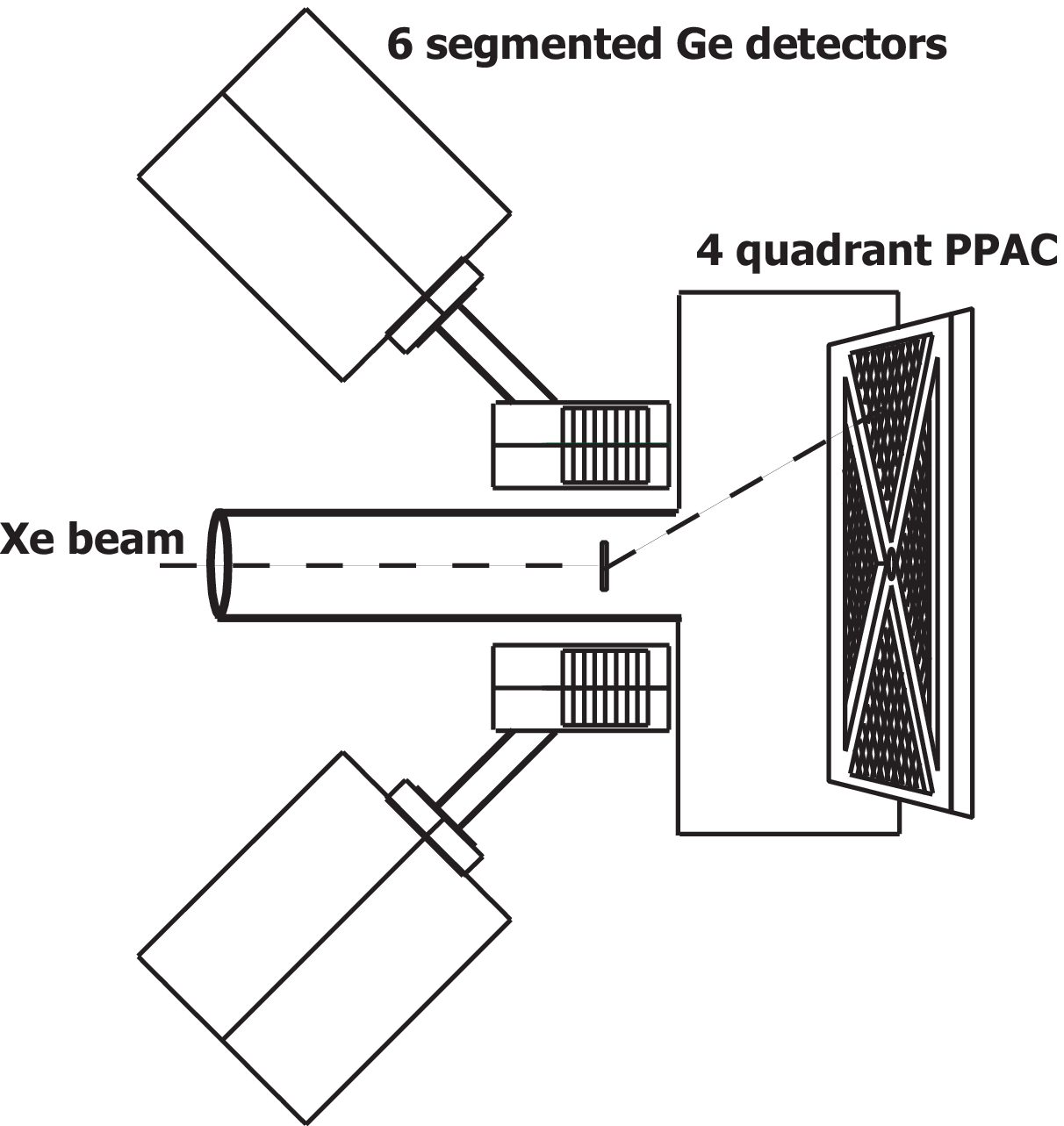}{5cm}{nlof}
{Schematic drawing of the
experimental setup used in the measurements.}

The $\gamma$ rays from the reaction were observed with a setup of
six germanium detectors from the MSU SeGA array~\cite{Mue01}.
These detectors were arranged in a barrel of 9.5~cm radius from
the middle of the target to the center of the germanium crystals.
In this arrangement, the array had an absolute photopeak
efficiency of 6.0\% at 1.3~MeV.  The germanium detectors of SeGA
are 32-fold segmented~\cite{Mue01} and allow for a precise Doppler
reconstruction of $\gamma$ rays emitted by nuclei in flight. The
total energy and timing information from each germanium detector
were obtained from the central contact. The efficiency of the
entire germanium array was determined with standard $\gamma$-ray
calibration sources covering an energy range from 88 to 1836 keV.
Corrections to this efficiency due to the summing of coincident
events in individual detectors were estimated with the use of {\sc
geant} simulations~\cite{gearef} of the setup.

For each event, the arrival of a scattered particle in the {\sc
ppac} was measured with respect to the {\sc atlas} resonators
frequency (rf).  This allows for a clear separation of the
small-angle ($<$90$^\circ$) and large-angle ($>$90$^\circ$)
scattering in the center-of-mass due to the significant difference
in the momentum (and consequently in time of flight) of the xenon
nuclei in the two cases.  Representative $\gamma$-ray spectra,
gated on a partial angle range of forward (in the center-of-mass)
scattered $^{128}$Xe nuclei, are shown in Fig.~\ref{fg:xespec}.

\wfmfig{xespec}{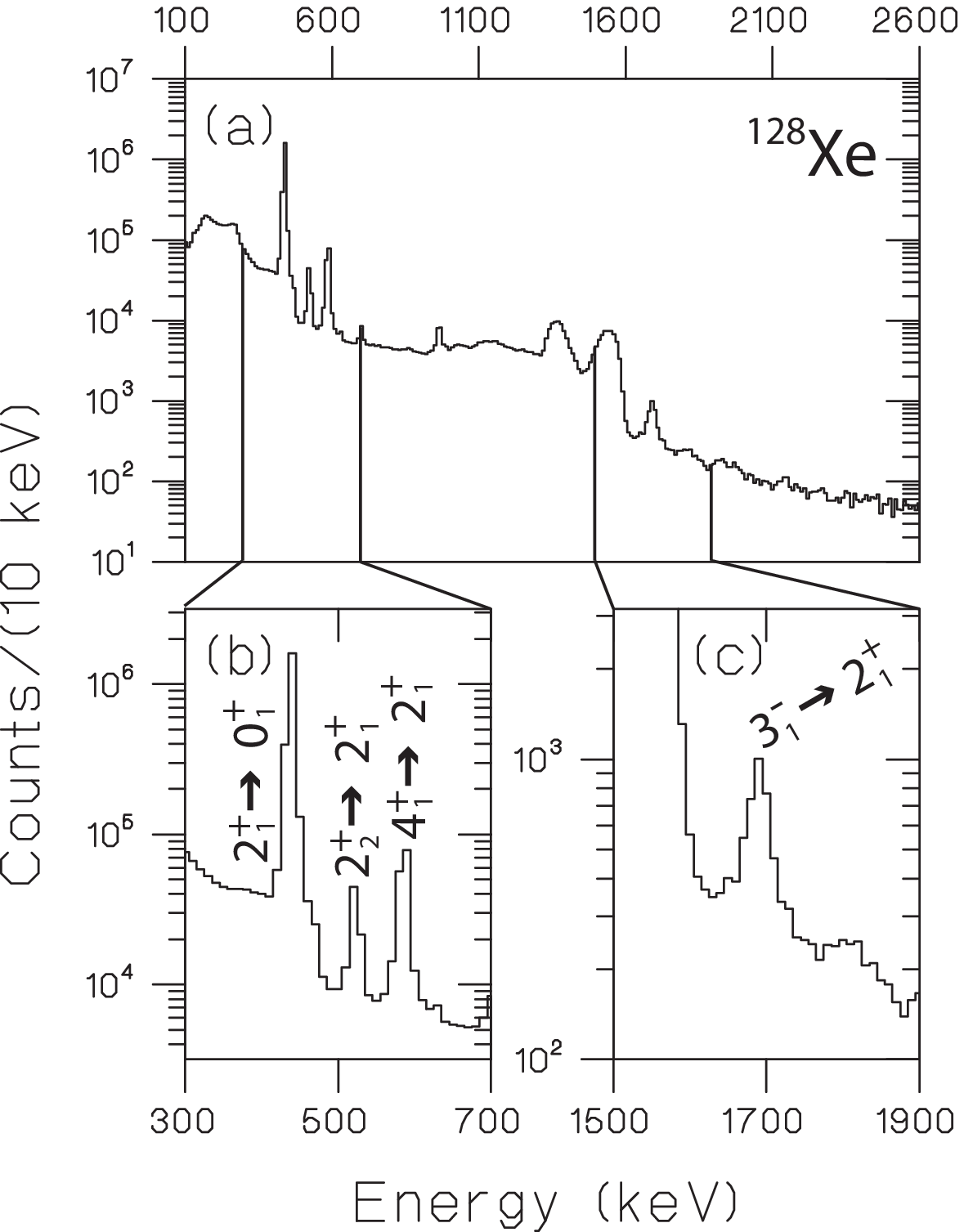}{7cm}{nlof}{Representative $\gamma$-ray
spectra for the $^{128}$Xe + $^{58}$Ni reaction in coincidence
with scattered $^{128}$Xe.  The spectra are gated on forward
scattering angles from 17.8$^\circ$ to 19.0$^\circ$ in the
laboratory frame. Panel (a) represents the energy range of
interest.  Panel (b) shows peaks corresponding to the $2^+_1
\rightarrow 0^+_1$, $2^+_2 \rightarrow 2^+_1$, and $4^+_1
\rightarrow 2^+_1$ $\gamma$-ray transitions. Panel  (c) highlights
the $3^-_1 \rightarrow 2^+_1$ transition of interest in the
present work.}

For the analysis of data on $^{124}$Xe and $^{126}$Xe, the
$\gamma$-ray events were sorted into spectra corresponding to five
equal divisions of scattering  angles between 10.6$^{\circ}$ and
16.0$^{\circ}$ in the laboratory frame. For the heavier Xe
isotopes, the five angle cuts were between 12.7$^{\circ}$ and
19.0$^{\circ}$. It should be noted that in all cases the condition
$r_{min} > (1.25(A_t^{1/3} + A_p^{1/3}) + 5)$~[fm] was satisfied,
excluding nuclear contributions to the excitation
process~\cite{Cli86}.

The $\gamma$-ray intensities corrected for the efficiency were
determined for each scattering angle. From these, level schemes
were constructed by comparing the measured energies to previously
known data~\cite{nds24,nds26,nds28,nds30,nds32,nds34} with
verification that the relative intensities were consistent with
Coulomb excitation. In cases where new $\gamma$ rays were
identified, placements are proposed  based on $\gamma$-$\gamma$
coincidences.

\wfmfig{xescheme}{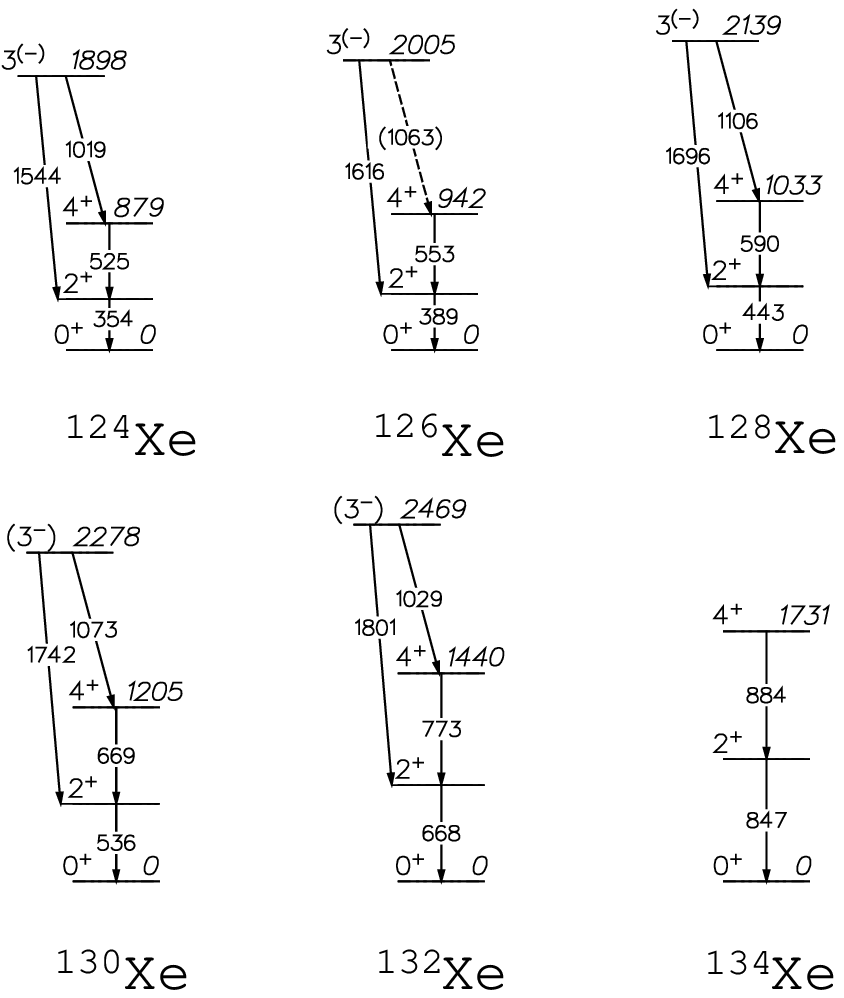}{8.2cm}{nlof}{Partial decay
schemes for even $^{124-134}$Xe isotopes.  Shown for each isotope
are the experimental (Exp.) 0$_1^+$, 2$_1^+$, and 4$_1^+$ levels,
as well as the proposed 3$_1^-$ state with corresponding
$\gamma$-ray transitions (dashed transitions correspond to known
$\gamma$ rays that are not observed in the present measurements).}

Excited states with confirmed spins of 3$\hbar$ have been clearly
identified in $^{124}$Xe, $^{126}$Xe, and $^{128}$Xe at
1898.0~\cite{Wer01}, 2004.8~\cite{Sei93}, and 2138.7
keV~\cite{Neu96}, respectively.  While the negative parity for
these states is suggested, there is circumstantial evidence to
support such an assignment in each case.

The observation of a 3$^-$ state in $^{130}$Xe has not been
reported previously. In the present experiment, several candidate
$\gamma$-ray transitions were observed.  Five transitions were
found to feed the 2$_1^+$ state in $^{130}$Xe and can be viewed as
candidates for decays from a 3$^-$ state.  Two of these $\gamma$
rays at 1687 and 1707 keV have been previously observed to feed
the first 2$^+$ level following $^{130}$Cs electron capture
decay~\cite{Hop73}.  However, the reported log$ft$ values are not
consistent with the forbidden decay that would be expected between
the 1$^+$ ground state of $^{130}$Cs and a 3$^-$ state. Of the
three remaining $\gamma$ rays with respective energies of 1742,
1896, and 2033 keV, the most intense---after efficiency
corrections---is the 1742 keV transition.  For $^{124,126,128}$Xe,
the $3^- \rightarrow 2^+$ transition was the dominant $\gamma$ ray
in the spectrum above 1400 keV (see e.g.\ Fig.~\ref{fg:xespec}).
Consequently, the 1742 keV transition is identified as a candidate
for the decay of the 3$_1^-$ state.  Based on sum-energy and
intensity-balance considerations, an observed 1073 keV $\gamma$
ray is placed feeding the 4$_1^+$ state from this same level. It
should be noted that, if this assignment is in error, the $B(E3;
0_1^+ \rightarrow 3_1^-)$ strength would be less than the value
presented below.

A candidate 3$^-$ state in $^{132}$Xe was proposed at an
excitation energy of 2469 keV by Gelletly~{\it et
al.}~\cite{Gel71}, from the decay by a 1801 keV $\gamma$ ray that
is also observed in the present experiment.  Unlike the lighter Xe
nuclei, where the strongest decay branch from the 3$^-$ level
occurs to the 2$_1^+$ state, the 2469 keV level in $^{132}$Xe has
rather significant de-excitation branches to the 4$_1^+$, 2$_2^+$,
and 2$_3^+$ states.

In $^{134}$Xe, a candidate for the 3$^-$ level has not yet been
identified.  In addition, none of the $\gamma$ rays identified in
the present experiment are consistent with expectations for a
3$^-$ state (i.e., a $\gamma$ ray feeding the 2$^+_1$ state with
an energy between 1500 to 2000 keV).  Consequently, an energy for
the 3$^-$ state is not proposed for $^{134}$Xe.  An upper limit on
the $B(E3;0^+_1 \rightarrow 3^-_1)$ strength was determined
assuming that the excitation energy for the 3$^-$ state is between
2400 and 3000 keV.

For the determination of the electromagnetic transition matrix
elements, the angle-dependent $\gamma$-ray yields were analyzed
using the Coulomb excitation code {\sc gosia}~\cite{Czo86,Czo91}.
This code combines the semi-classical theory of multiple Coulomb
excitation~\cite{Ald75} and the measured $\gamma$-ray de-excitation
patterns with a numerical least-squares analysis to determine the
electromagnetic matrix elements from the experimental $\gamma$-ray
yields. In addition to the $\gamma$-ray yields determined in the
present experiments, the previously observed $\gamma$-ray
branching ratios and multipole mixing ratios for relevant
transitions were used when possible as additional information to
constrain the determination of the matrix elements further. For
those cases where a branching ratio is not known (e.g.~the 3$^-$
state in $^{130}$Xe), the ratio is fit based solely on the
presently measured data.

All observed $\gamma$-ray yields were used in the calculation of
the transition matrix elements in each of the Xe nuclei.  This
includes transitions not presented in Fig.~\ref{fg:xescheme}. 
$\gamma$-ray branching ratios for transitions from the 3$^-$ states 
are listed in Table~\ref{tab:branching}. 
The
determination of the $M(E3)$ transition matrix elements was
carried out using the method described in Ref.~\cite{Bur85}.
Specifically, the $M(E1)$ transition matrix elements (e.g.\ $M(E1;
2_1^+ \rightarrow 3^-)$ and $M(E1; 4_1^+ \rightarrow 3^-)$) were
set equal to a total strength of 10$^{-4}$ W.u., with relative
strengths adjusted to reproduce the branching ratios (if known).
$E1$ strengths of 10$^{-4}$ W.u. are typical for this mass region
(see e.g.\ Ref.~\cite{Cot93}).  The static quadrupole moments for
the 3$^-$ states ($Q(3^-)$) were set to zero.  The $M(E3; 0_1^+
\rightarrow 3^-)$ matrix element (as well as the matrix elements
of other transitions) were then varied to reproduce the observed
yields as a function of the xenon scattering angle for the $3^-
\rightarrow 2_1^+$ and other $\gamma$-ray transitions seen in this
work. To check the sensitivity of the calculation with respect to
correlations between the extracted $E1$ and $E3$ strengths, the $M(E1)$
parameters were increased and decreased by as much as a factor of ten,
and in all cases the change of the extracted $M(E3)$ matrix element was
found to be less than 2.5\%. This indicates, that in the excitation
strength of the $3^-_1$ state, the one-step $E3$ strength dominates over
two-step excitations, leading to a high experimental sensitivity for
$B(E3)$ strength. Additionally, variations of the $Q(3^-)$
moment by $\pm 0.5$ $e$b resulted in changes in the extracted values
of less than $\pm 5$\% for all cases.  These 
uncertainties are included in the overall errors of the
experimental results quoted in this work.

\begin{table}
\caption[]{$\gamma$-ray branching ratios for transitions from the 3$^-$ states for different Xe isotopes.}
\label{tab:branching}
\begin{ruledtabular}
\begin{tabular}{lcccc}
          &E(3$^-$) & E$_f$ & E$_{\gamma}$ &Branching\\
          &(keV)    &(keV)  & (keV)        &ratio\\
\hline
$^{124}$Xe &1898$^*$     &354    &1544          &100(13)\\
           &         &879    &1019$^{\dag}$          &16(8)\\
$^{126}$Xe &2005     &389    &1616          &100(17)\\
           &         &942    &1063          &21(1)\\
$^{128}$Xe &2139     &443    &1696          &100(3)\\
           &         &1033   &1106          &\\
$^{130}$Xe &2278     &536    &1742$^{\dag}$          &54(10)\\
           &         &1205   &1073$^{\dag}$          &100(26)\\
$^{132}$Xe &2469     &668    &1801          &56(11)\\
           &         &1298   &1171          &11(2)\\
           &         &1440   &1029          &69(8)\\
           &         &1963   &506           &$<$6.4\\
           &         &1986   &483           &100(9)\\
           &         &2040   &429           &8(2)\\
\end{tabular}
\end{ruledtabular}
\begin{flushleft}
$^*$ Negative parity proposed by the present work.\\
$^{\dag}$ $\gamma$-ray not previously reported.
\end{flushleft}
\end{table}

\begin{table}
\caption[]{Reduced transitions rates for $A=124-136$ even Xe
nuclei. Those columns listed ``present'' are the transition rates
from the measurements reported here (in both e$^2$b$^\lambda$ and
Weisskopf units), while ``previous'' are $B(E2 \uparrow)$ rates
from the references listed in square brackets.}  \label{tab:xebe2}
\begin{ruledtabular}
\begin{tabular}{lcccccc}
           & \multicolumn{4}{c}{$B(E2; 0^+_1 \rightarrow 2^+_1)$} & \multicolumn{2}{c}{$B(E3; 0^+_1 \rightarrow 3^-_1)$} \\
           & \multicolumn{2}{c}{present} & \multicolumn{2}{c}{previous} & \multicolumn{2}{c}{present} \\
           & (e$^2$b$^2$) & (W.u.) & \multicolumn{2}{c}{(W.u.)} & (e$^2$b$^3$) & (W.u.) \\
\hline
$^{124}$Xe & 1.12$^{+12}_{-9}$   & 60.8$^{+62}_{-50}$ & 57.8$^{+15}_{-14}$  & \cite{sah04} & 0.091(10) & 14.3(16)\\
$^{126}$Xe & 1.02$^{+13}_{-6}$   & 54.2$^{+70}_{-30}$ & 40(14)   & \cite{nds26} & 0.085(13) & 13(2)\\
$^{128}$Xe & 0.825$^{+11}_{-12}$ & 42.6$^{+54}_{-64}$ & 40.2(55) & \cite{nds28} & 0.083(11) & 12(2)\\
$^{130}$Xe & 0.585$^{+9}_{-6}$   & 30.0$^{+44}_{-28}$ & 37.2(17) & \cite{Jak02} & 0.033(9) & 4.7(13)\\
$^{132}$Xe & 0.499$^{+36}_{-32}$ & 25.0$^{+18}_{-16}$ & 23.7(6)  & \cite{Jak02} & 0.016(6)  & 2.1(9)\\
$^{134}$Xe & 0.322$^{+41}_{-16}$ & 15.8$^{+20}_{-8}$  & 14.7(1)  & \cite{Jak02} & $<$0.011  & $<$1.5\\
\end{tabular}
\end{ruledtabular}
\end{table}

The deduced $B(E2; 0_1^+ \rightarrow 2_1^+)$ and $B(E3; 0_1^+
\rightarrow 3_1^-)$ rates are presented in Table~\ref{tab:xebe2}.
Also given are the $B(E2; 0_1^+ \rightarrow 2_1^+)$ values from
the references listed in the table.  There is good general
agreement between the previously measured $B(E2; 0_1^+ \rightarrow
2_1^+)$ values and those determined in the present work.

Figure~\ref{fg:ene_be3} shows the experimental and calculated
excitation energies $E(3^-_1)$ and $B(E3 \uparrow)$ strengths. It
can be seen that the excitation energy of the $3^-_1$ state
increases smoothly from 1898 keV in $^{124}$Xe to 2469 keV in
$^{132}$Xe with the $B(E3 \uparrow)$ strength gradually decreasing
towards $^{134}$Xe.  As discussed in \cite{Met95}, collective
octupole motion in such nuclei is strongly influenced by
quadrupole softness. Qualitatively speaking, the octupole motion
proceeds on the background of a slowly changing dynamic quadrupole
deformation. Simple semi-microscopic arguments predict the energy
of the octupole phonon to be correlated with that of the
quadrupole phonon as
\begin{equation}
E(3^-)=E_0 - \frac{B^2}{E(2^+_1)}.
\end{equation}
In Fig.~\ref{fg:ene_be3}, our experimental results are compared to
eq. (1), where $B=0.7$~MeV, $E_0=3.275$~MeV (from the energy of the 3$^-_1$ state in the closed-shell nucleus $^{136}$Xe) and $E(2^+_1)$ is in MeV.

\wfmfig{ene_be3}{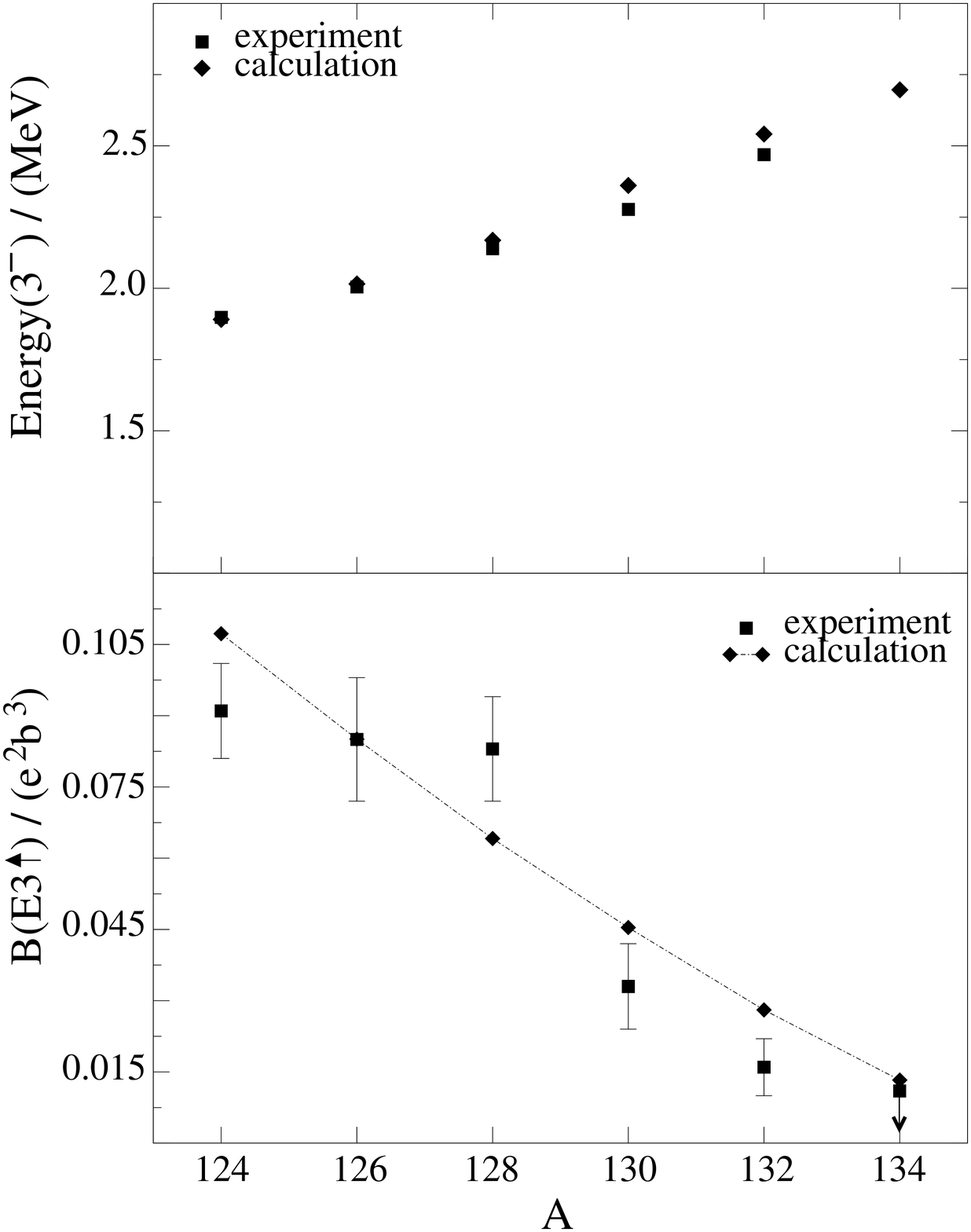}{8.4cm}{nlof}{Experimental results
  compared to phenomenological calculations outlined in the text.}

The global behavior of the octupole strength throughout the
nuclear chart was described in \cite{Met95} by the expression of
the type
\begin{equation}
B(E3 \uparrow) = \kappa Z^2 A^{1/3}\frac{1}{E(3^-)} \xi
~~~\rm{[e^2b^3]}
\end{equation}
using the universal $Z$- and $A$-dependence, the parameter
$\kappa$ setting the global scale and the local parameter $\xi$
depending on pairing and details of the shell structure. For xenon
nuclei, the $3^-_1$ state is likely dominated by $\Delta j =
\Delta l = 3$, $d_{5/2} \otimes h_{11/2}$ correlations.
The occupation of the $h_{11/2}$ orbital gradually increases for
heavier xenon isotopes until $^{136}$Xe, where the $\nu h_{11/2}$
configuration is completely filled $(N=82)$ and thus octupole
correlations involving this orbit are blocked. Using $\kappa=1.4
\cdot 10^{-5}~e^2$b$^3$, the value close to the average value
chosen in \cite{Met95}, $E(3^-)$ in MeV, and the occupation factor
$\xi=(82-N)/12$ as a measure of the availability of the $\nu
h_{11/2}$ orbit to form octupole correlations, we come to the fit
of Fig. 4. Detailed microscopic calculations would require a theory
taking into account strong quadrupole anharmonicity, 
coupling between the modes beyond the conventional random phase
approximation, and possible fragmentation of the octupole
strength.

The $N=82$ nucleus $^{136}$Xe was not a subject of the present
study. The first $3^-$ state has been 
established at 3275~keV and a $B(E3)$ value of 17(5)~W.u. was determined
in inelastic proton scattering \cite{Sen72}. The octupole strength
observed in this semi-magic nucleus is not related to the octupole correlations
formed with participation of the $h_{11/2}$ orbit and is thus outside our
phenomenological description of the properties of the lighter Xe
isotopes. The growth of octupole strength is also known for other
magic nuclei, for example for $^{40}$Ca~\cite{Kib02}.

The data presented here represent the first determination of the
$B(E3; 0_1^+ \rightarrow 3_1^-)$ values in a chain of even-even
$^{124-134}$Xe nuclei. The trends in energy $E(3^-)$ and $B(E3;
0_1^+ \rightarrow 3_1^-)$ excitation strengths are well reproduced
using phenomenological models with the universal $A$- and
$Z$-dependence. The smoothly increasing excitation energy of the
3$^-_1$ states from $^{124}$Xe to $^{134}$Xe and the decreasing
$B(E3 \uparrow)$ strength can be understood, at least
qualitatively, as being associated with the presence of the soft
quadrupole mode of the mean field and the decreased availability of
the $\nu h_{11/2}$ orbital for the generation of octupole
correlations.

The authors thank the staff of {\sc atlas} for providing
high-quality xenon beams. W.F.~Mueller thanks D.~Cline and
C.~Y.~Wu for significant assistance in using the {\sc gosia} code
and acknowledges fruitful discussions with P.D.~Cottle and
R.M.~Ronningen. A.G. acknowledges support from Professor P. von
Brentano and the University of Cologne. 
This work was supported by the National Science
Foundation under grant numbers PHY-0110253, PHY-9875122,
PHY-0244453, and by the United States Department of Energy, Office
of Nuclear Physics, under contract numbers W-31-109-ENG-38 and
DE-FG02ER41220.



\begin{thebibliography}{}
\bibitem{Cas85} R.F.~Casten and P.~von~Brentano, Phys.\ Lett.\
  {\bf 152}, 22 (1985).
\bibitem{Zel90} V.G. Zelevinsky, In: {\sl New Trends in Nuclear
Collective Dynamics}, Genshikaku Kenkyu (Tokyo) {\bf 35}, 21
(1991); see also discussion in G.F. Bertsch, Nucl. Phys. {\bf
A574}, 169c (1994).
\bibitem{Neu96} U.~Neuneyer {\it et al.},
  Nucl.\ Phys. A {\bf 607}, 299 (1996).
\bibitem{Gad00} A.~Gade {\it et al.}, Nucl.\ Phys.\ A {\bf 665}, 268 (2000).
\bibitem{Wer01} V.~Werner {\it et al.}, 
  Nucl.\ Phys.\ A {\bf 692}, 451 (2001).
\bibitem{Pan96} X.-W.~Pan {\it et al.}, 
  Phys.\ Rev.\ C {\bf 53}, 715 (1996).
\bibitem{Mey98} U.~Meyer {\it et al.},
  Nucl.\ Phys.\ A {\bf 641}, 321 (1998).
\bibitem{Yos04} N.~Yoshinaga and K.~Higashiyama, Phys.\ Rev.\ C
  {\bf 69}, 054309 (2004).
\bibitem{Met95} M.P. Metlay, J.L. Johnson, J.D. Canterbury,
P.D. Cottle, C.W. Nestor, Jr., S. Raman and V.G. Zelevinsky, Phys.
Rev. C {\bf 52}, 1801 (1995).
\bibitem{Spe90} R.H.~Spear and W.N.~Catford, Phys.\ Rev.\ C
  {\bf 41}, 1351 (1990).
\bibitem{Naz84} W.~Nazarewicz {\it et al.}, 
  Nucl.\ Phys.\ A
  {\bf 429}, 269 (1984).
\bibitem{Jon81} N.-G.~Jonsson {\it et al.}, 
  Nucl.\ Phys.\ A {\bf 371}, 333 (1981).
\bibitem{Mat75} M.~Matoba {\it et al.}, 
  Nucl.\ Phys.\ A {\bf 237}, 260 (1975).
\bibitem{Bur85} S.~M.~Burnett {\it et al.}, 
  Nucl.\ Phys.\ A {\bf 432}, 514 (1985).
\bibitem{Sen72} S.~Sen {\it et al.}, 
  Phys.\ Rev.\ C
  {\bf 6}, 2201 (1972).
\bibitem{Kib02} T.~Kib\'edi and R.H.~Spear, At.\ Data \& Nucl.\ Data Tables
  {\bf 80}, 35 (2002).
\bibitem{Fla03} V.V.\ Flambaum and V.G.\ Zelevinsky. Phys. Rev. C {\bf 68},
035502 (2003).
\bibitem{Mue01} W.F.~Mueller {\it et al.}, 
  Nucl.\ Inst.\ and Meth.\ A {\bf 466}, 492 (2001).
\bibitem{gearef} GEANT-detector description and simulation tool,
  version 3.21, Technical Report W5013, CERN (1994).
\bibitem{Cli86} D.~Cline, Ann.\ Rev.\ Nucl.\ Part.\ Sci.\
  {\bf 36}, 683 (1986).
\bibitem{nds24} H.~Iimura {\it et al.}, 
  Nucl.\ Data Sheets {\bf 80}, 895 (1997).
\bibitem{nds26} J.~Katakura and K.~Kitao,
  Nucl.\ Data Sheets {\bf 97}, 765 (2002).
\bibitem{nds28} M.~Kanbe and K.~Kitao,
  Nucl.\ Data Sheets {\bf 94}, 227 (2001).
\bibitem{nds30} B.~Singh,
  Nucl.\ Data Sheets {\bf 93}, 33 (2001).
\bibitem{nds32} Yu.V.~Sergeenkov,
  Nucl.\ Data Sheets {\bf 65}, 277 (1992).
\bibitem{nds34} A.A.~Sonzogni,
  Nucl.\ Data Sheets {\bf 103}, 1 (2004).
\bibitem{Sei93} F.~Seiffert {\it et al.}, 
  Nucl.\ Phys.\ A {\bf 554}, 287 (1993).
\bibitem{Hop73} P.K.~Hopke {\it et al.}, 
  Phys.\ Rev.\ C {\bf 8}, 745 (1973).
\bibitem{Gel71} W.~Gelletly {\it et al.}, 
  Phys.\ Rev.\ C {\bf 3}, 1678 (1971).
\bibitem{Czo86} T.~Czosnyka {\it et al.}, 
  Nucl.\ Phys.\ A {\bf 458}, 123 (1986).
\bibitem{Czo91} T.~Czosnyka {\it et al.}, 
  {\sc gosia} users manual, NSRL-305, 1991.
\bibitem{Ald75} K.~Alder and A.~Winther,
  Electromagnetic Excitation Theory of Coulomb Excitation with Heavy Ions,
  North Holland, Amsterdam (1975).
\bibitem{sah04} B.~Saha {\it et al.}, Phys.\ Rev.\ C {\bf 70},
  034313 (2004).
\bibitem{Jak02} G.~Jakob {\it et al.}, Phys.\ Rev.\ C {\bf 65},
  024316 (2002).
\bibitem{Cot93} P.D.~Cottle, Phys.\ Rev.\ C {\bf 47}, 1529 (1993).

\end{thebibliography}
\end{document}